\documentclass{PoS}
\def \pt {p_{\rm T}}

\title{Production of jets accompanied by W and Z bosons at LHC startup}

\ShortTitle{Production of jets accompanied by W and Z bosons}

\author{\speaker{Didar Dobur}\thanks{On behalf of CMS Collaboration}\\
        Istituto Nucleare di Fisica Nazionale(INFN) Pisa, Italy\footnote{Now at the University of Florida, Gainsville, USA}.\\
        E-mail: \email{ddidar@mail.cern.ch}}

\abstract{We report on potential for measurement of W and Z boson production accompanied by jets. 
Of particular interest are jet multiplicity and $P_{\rm T}$ distributions. The 10 to 100 $pb^{-1}$ datasets 
expected in the startup year of operation of LHC are likely to already provide information beyond 
the reach of the Tevatron collider both in jet multiplicity and $P_{\rm T}$ range. We are especially interested 
in understanding the ratios of W+jets to Z+jets distributions by comparing them to next-to-leading 
order Monte Carlo generators, as these processes present a formidable background for searches of new 
physics phenomena.}

\FullConference{2008 Physics at LHC\\
		 September 29 - 4 October 2008\\
		 Split, Croatia}

\begin{document}

\section{Introduction}
The production of W(Z) bosons associated with jets at LHC has a wide range of physics potential, which varies 
from Standard Model (SM) measurements to Supersymmetry (SUSY) searches. 
These processes can be used for tests of perturbative quantum chromodynamic (QCD)~\cite{QCD}. 
The predictions for W(Z)$+N$Jets, where $N>2$, are accessible only 
through matrix element (ME) plus parton shower (PS) computations and in fact, 
can be considered as a prime testing ground for the accuracy of such predictions.
Z+jet events can be also exploited to calibrate jets, measured in the Calorimeter 
(See~\cite{jetmet} for details). 
Furthermore, W(Z)+jets form a relevant background to many interesting phenomena, including new physics. Therefore, 
these processes must be measured with great accuracy to allow precision measurements and 
increase the sensitivity of the searches beyond SM.

However, the individual cross section measurements
of W+Njets and Z+Njets will be affected by large systematic 
uncertainties associated mostly with the definition and measurement of jets. 
One of the measurements that CMS plans to perform is the ratio of the cross sections of W+jets to Z+jets 
as functions of jet multiplicity and boson $\pt$. 
Such a measurement allows partial cancellation 
of the most relevant experimental systematic uncertainties as well as the theoretical uncertainties due to
the choice of renormalization scale, the parton distribution functions, etc.~\cite{wzjets}. 
The jet energy scale forms the largest 
experimental uncertainty as it increases rapidly with jet multiplicity. This uncertainty cancels in the ratio as long as 
the $\pt$ spectra, the rapidity distribution and the 
composition of the jets in both processes are the same at the level of experimental sensitivity. 
Other uncertainties, i.e. Underlying Event (UE), Multiple interactions, 
luminosity and detector acceptances, will also cancel to a large
extend in the ratio.  
%

A number of physics generators are available to simulate major kinematic 
properties of \\ 
W(Z)+jets. The measurements of W(Z)+jets at the Tevatron collider 
indicate a general agreement between the theoretical predictions based on 
LO ME plus PS and data~\cite{CDF}.
In the studies presented here the ME event generator ALPGEN 
is used to generate exclusive parton level W(Z)+Njets (N=0,1,2,3,4,5) events. 
PYTHIA is used for PS and hadronization.   
The MLM recipe is used in order to avoid double counting of processes from ME and PS.  
The SM processes $t\bar{t}$+jets, WW+jets, WZ+jets, ZZ+jets and QCD
multi-jet are considered as backgrounds and generated with PYTHIA 
in fully inclusive decay modes for W and Z bosons. 

\begin{figure}
\begin{center}
  \includegraphics[scale=0.60]{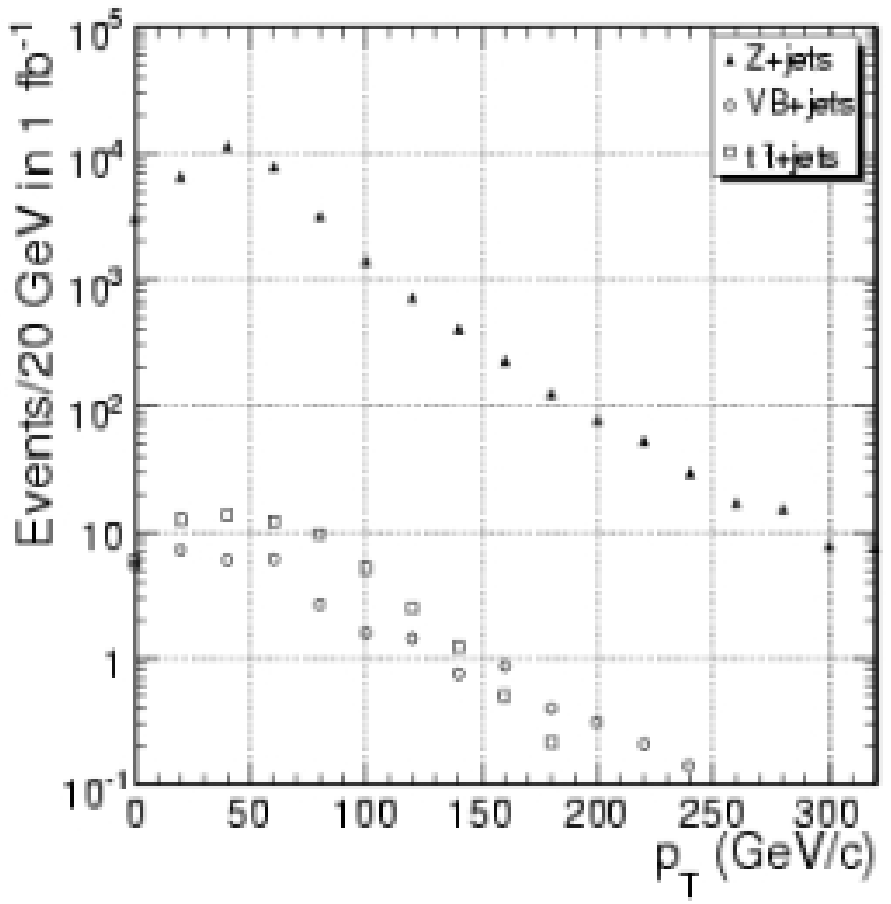}
  \includegraphics[scale=0.60]{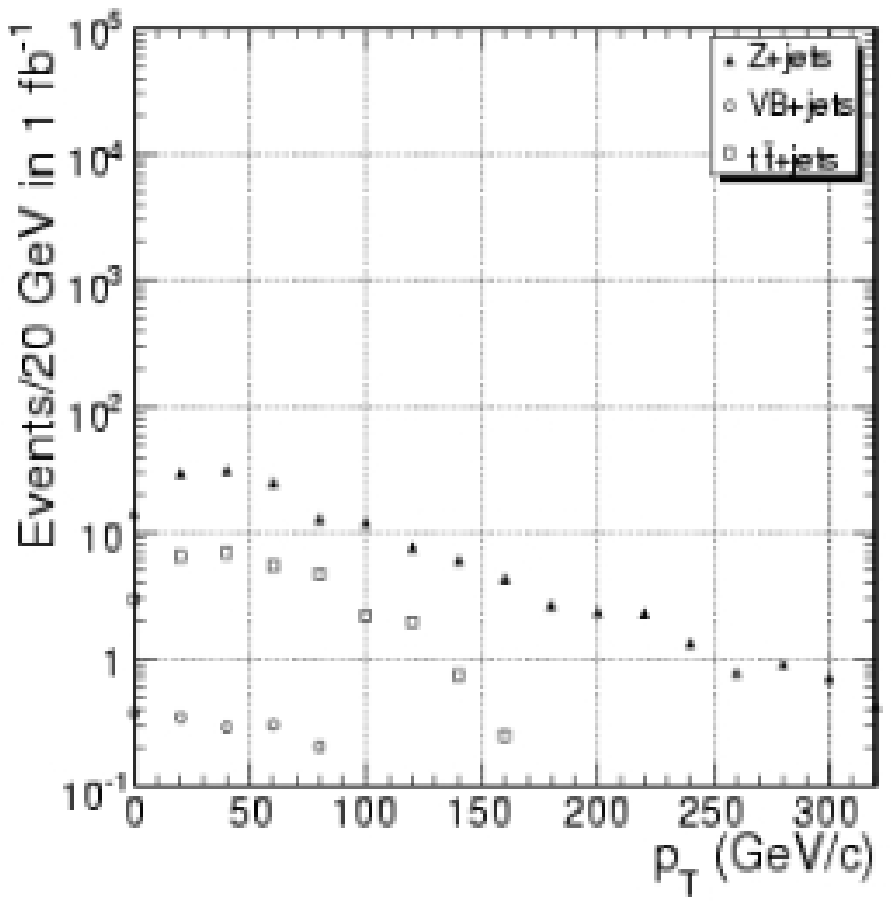} 
  \caption{$\pt$ distribution of the Z boson in selected Z+$\ge 1$jet (left) and Z+$\ge 4$jet 
  (right) for signal and background for an integrated luminosity of $1~fb^{-1}$.}
  \label{fig:zjets}
\end{center}
\end{figure}

\begin{figure}
\begin{center}
  \includegraphics[scale=0.60]{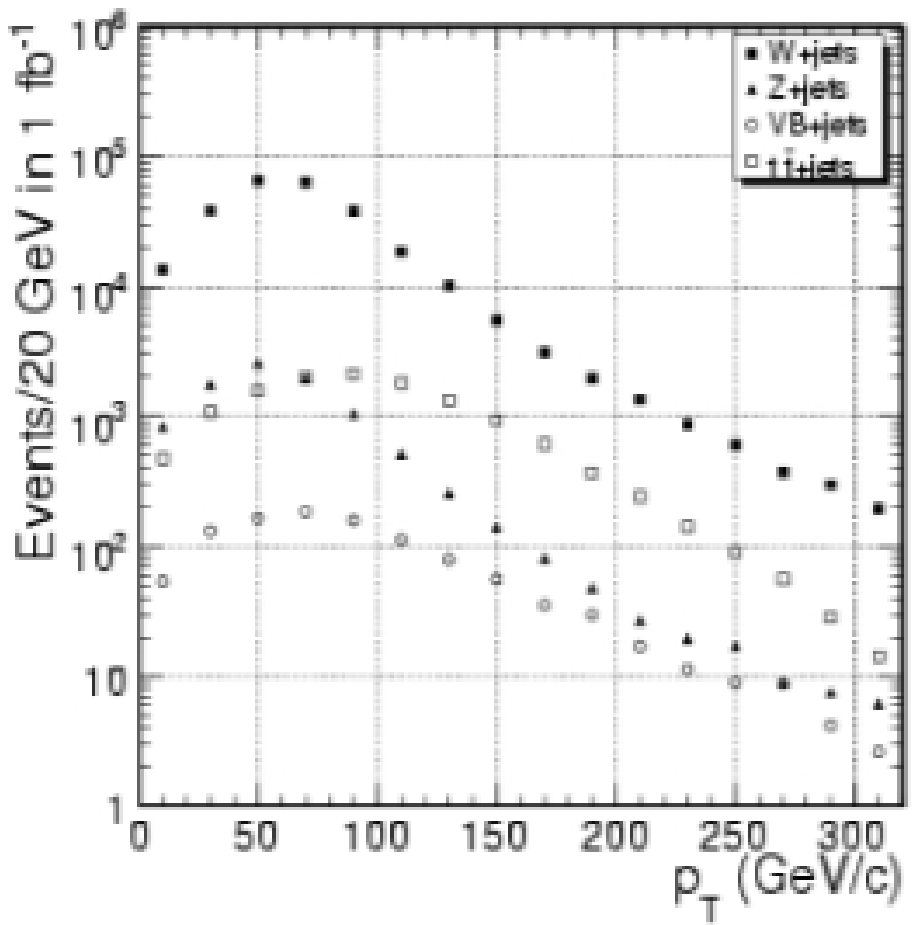}
  \includegraphics[scale=0.60]{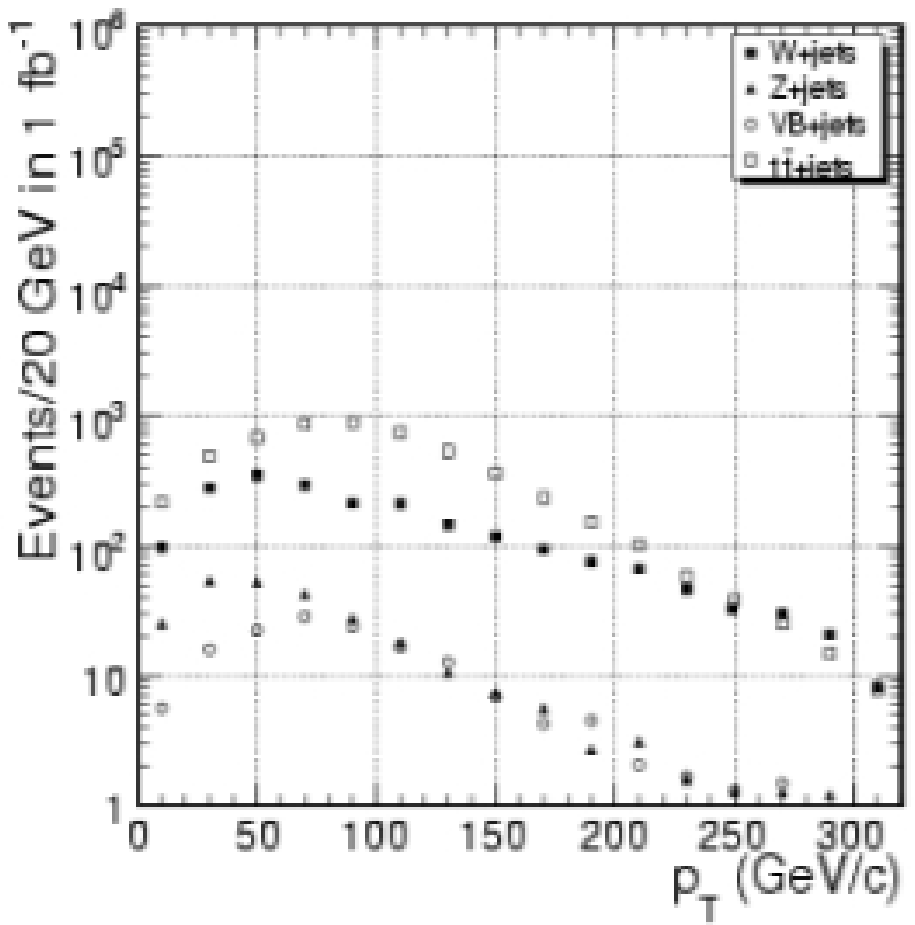} 
  \caption{$\pt$ distribution of the W boson in selected W+$\ge 1$jet (left) and W+$\ge 4$jet 
  (right) for signal and background for an integrated luminosity of $1~fb^{-1}$.}
  \label{fig:wjets}
\end{center}
\end{figure}

Figure~\ref{fig:zjets}(\ref{fig:wjets}) shows the $\pt$ distribution of the Z(W) boson in selected 
Z(W)+$\ge 1$jet (left) and Z(W)+$\ge 4$jet (right) events for the signal and backgrounds. In both W and Z boson cases 
the events are selected in  the electron and muon channels. The high $\pt$ isolated leptons are selected in 
order to reduce
contamination from QCD events. Furthermore, Z+jets events are selected by a tight di-lepton invariant mass around the
Z boson mass and the Missing transverse energy is restricted to be small, whereas for W+jets a large missing transverse
energy is required. Jet reconstruction is performed using the Iterative Cone algorithm
using the energy deposited in the Calorimeter. Jets are calibrated using $\gamma$+jet events and the jets with 
$\pt > 50~ {\rm GeV}$ are counted. The current Z+jets selection provides a rather "clean" sample, and with 
$1~fb^{-1}$ of data, up to fourth jet multiplicity can be measured.   
One crucial point will be the reduction of the background to the W+>2jets from $t\bar{t}$ events 
(See Fig.~\ref{fig:wjets} right), since the 
$t\bar{t}$ production rate increases by about a factor of 100 from the Tevatron to the LHC, while 
W production increases by just a factor of 5.  In the studies presented, the QCD contribution as 
background to W(Z)+jets is found to be negligible
and not shown in the figures. However, we should note that the background processes are 
simulated using the PYTHIA program,
which is known not to produce high jet multiplicities correctly.
The data driven methods for background estimation are therefore
extremely important.

The cross section measurements of W(Z)+jets versus 
the jet multiplicity will be one of the early measurements carried out with the CMS detector. 
The ratio measurement of W+Njets to Z+Njets will allow partial cancellation of the most relevant 
systematic uncertainties. This is an extremely important advantage at the startup, where it will be 
relatively difficult to control the
systematic uncertainties. The ratio measurement can benefit from other types of jets 
(e.g. reconstructed with Tracker Tracks only) than the standard Calorimeter
based jets. Unlike Calorimeter jets Track based jets preserve the vertex information which helps to count jets originating only 
from the signal vertex, eliminating Pile Up contamination to the jets.  

\end{document}